\documentclass[aps,prl,showpacs,floatfix,twocolumn,byrevtex,superscriptaddress]{revtex4-1}
\usepackage{amsmath}
\usepackage{amssymb}
\usepackage{amstext}
\usepackage{amsopn}
\usepackage{amsfonts}
\usepackage{amsxtra}
\usepackage[english]{babel}
\usepackage{graphicx}
\usepackage{float}
\usepackage{bm}
\usepackage{multirow}
\usepackage{dcolumn}
\usepackage{color}
\usepackage{hyperref}

\newcommand{\icm}{\ensuremath{\textrm{cm}^{-1}}}

\newcommand{\CVS}{CsV$_{3}$Sb$_{5}$}

\newcommand{\AVS}{$A$V$_{3}$Sb$_{5}$}

\newcommand{\CVNS}{Cs(V$_{1-x}$Nb$_{x}$)$_{3}$Sb$_{5}$}
\newcommand{\EF}{$E_{\text{F}}$}

\begin{document}

\title{Effects of Nb Doping on the Charge-Density Wave and Electronic Correlations in the Kagome Metal Cs(V$_{1-x}$Nb$_{x}$)$_{3}$Sb$_{5}$}
\author{Xiaoxiang~Zhou}
\thanks{These authors contributed equally to this work.}
\affiliation{National Laboratory of Solid State Microstructures and Department of Physics, Collaborative Innovation Center of Advanced Microstructures, Nanjing University, Nanjing 210093, China}
\author{Yongkai~Li}
\thanks{These authors contributed equally to this work.}
\affiliation{Centre for Quantum Physics, Key Laboratory of Advanced Optoelectronic Quantum Architecture and Measurement (MOE), School of Physics, Beijing Institute of Technology, Beijing 100081, China}
\affiliation{Beijing Key Lab of Nanophotonics and Ultrafine Optoelectronic Systems, Beijing Institute of Technology, Beijing 100081, China}
\affiliation{Material Science Center, Yangtze Delta Region Academy of Beijing Institute of Technology, Jiaxing 314011, China}
\author{Zhe~Liu}
\thanks{These authors contributed equally to this work.}
\author{Jiahao~Hao}
\author{Yaomin~Dai}
\email{ymdai@nju.edu.cn}
\affiliation{National Laboratory of Solid State Microstructures and Department of Physics, Collaborative Innovation Center of Advanced Microstructures, Nanjing University, Nanjing 210093, China}
\author{Zhiwei~Wang}
\email{zhiweiwang@bit.edu.cn}
\affiliation{Centre for Quantum Physics, Key Laboratory of Advanced Optoelectronic Quantum Architecture and Measurement (MOE), School of Physics, Beijing Institute of Technology, Beijing 100081, China}
\affiliation{Beijing Key Lab of Nanophotonics and Ultrafine Optoelectronic Systems, Beijing Institute of Technology, Beijing 100081, China}
\affiliation{Material Science Center, Yangtze Delta Region Academy of Beijing Institute of Technology, Jiaxing 314011, China}
\author{Yugui~Yao}
\affiliation{Centre for Quantum Physics, Key Laboratory of Advanced Optoelectronic Quantum Architecture and Measurement (MOE), School of Physics, Beijing Institute of Technology, Beijing 100081, China}
\affiliation{Beijing Key Lab of Nanophotonics and Ultrafine Optoelectronic Systems, Beijing Institute of Technology, Beijing 100081, China}
\author{Hai-Hu~Wen}
\email{hhwen@nju.edu.cn}
\affiliation{National Laboratory of Solid State Microstructures and Department of Physics, Collaborative Innovation Center of Advanced Microstructures, Nanjing University, Nanjing 210093, China}

\date{\today}
%
%

\begin{abstract}
The transport and optical properties of the Nb-doped Cs(V$_{1-x}$Nb$_{x}$)$_{3}$Sb$_{5}$ with $x = 0.03$ and 0.07 have been investigated and compared with those of the undoped CsV$_{3}$Sb$_{5}$. Upon Nb doping, the charge-density wave (CDW) transition temperature $T_{\text{CDW}}$ is suppressed, and the superconducting temperature $T_{c}$ rises. The residual resistivity ratio decreases with Nb doping, suggesting an increase of disorder. For all compounds, the optical conductivity in the pristine phase reveals two Drude components (D1 and D2). The substitution of Nb causes an increase of D1 alongside a reduction of D2 in weight, which implies a change of the Fermi surface. The total Drude weight is reduced with increasing Nb content, signifying an enhancement of electronic correlations. Below $T_{\text{CDW}}$, while the optical conductivity clearly manifests the CDW gap in all materials, the gapped portion of the Fermi surface shrinks as the Nb content grows. A comprehensive analysis indicates that the change of the Fermi surface, the enhancement of electronic correlations, the shrinkage of the removed Fermi surface by the CDW gap, and the increase of disorder may all have a considerable impact on the interplay between the CDW and superconductivity in Cs(V$_{1-x}$Nb$_{x}$)$_{3}$Sb$_{5}$.
\end{abstract}

\maketitle

%
%
The kagome lattice has generated substantial interest, as its electronic structure consists of flat bands, Dirac points and saddle points, which support a myriad of exotic quantum phenomena~\cite{Ko2009PRB,Guo2009PRB,Yu2012PRB,Kiesel2012PRB,Kiesel2013PRL,Wang2013PRB,Ferrari2022PRB}. When the Fermi level \EF\ is situated near the saddle point, i.e. at the van Hove filling, the ground state of the kagome lattice system sensitively depends on the local $U$ and nearest-neighbor $V$ Coulomb interactions~\cite{Yu2012PRB,Kiesel2012PRB,Kiesel2013PRL,Wang2013PRB,Denner2021PRL,Wu2021PRL,Ferrari2022PRB}, resulting in a rich phase diagram that comprises charge or spin bond order~\cite{Kiesel2013PRL,Wang2013PRB,Denner2021PRL}, unconventional superconductivity~\cite{Yu2012PRB,Kiesel2012PRB,Kiesel2013PRL,Wang2013PRB,Wu2021PRL}, ferromagnetism~\cite{Kiesel2013PRL,Wang2013PRB}, charge-density wave (CDW)~\cite{Wang2013PRB,Denner2021PRL,Ferrari2022PRB} and spin-density wave (SDW)~\cite{Yu2012PRB}.

Recently, a new family of kagome metals \AVS\ ($A$ = K, Rb, Cs) with an ideal two-dimensional kagome network of V has been discovered~\cite{Ortiz2019PRM} and, moreover, multiple saddle points formed by V-$3d$ orbitals have been identified near \EF~\cite{Ortiz2019PRM,Ortiz2020PRL,Park2021PRB,LaBollita2021PRB,Tan2021PRL,Nakayama2021PRB,Cho2021PRL,Liu2021PRX,Hu2022NC,Kang2022NP,Nakayama2022PRX}, setting the stage for the investigation of the theoretically predicted exotic quantum states at the van Hove filling. These kagome metals undergo a CDW transition at $T_{\text{CDW}}$ = 78, 103, and 94~K, respectively, before entering a superconducting phase below $T_{\text{c}}$ = 0.93, 0.92, and 2.5~K, respectively~\cite{Ortiz2020PRL,Ortiz2021PRM,Yin2021CPL}. The suppression of the CDW order via pressure~\cite{Yu2021NC,Chen2021PRL,Du2021PRB,Zhang2021PRB}, uniaxial strain~\cite{Qian2021PRB}, or chemical doping~\cite{Song2021PRL,Oey2022PRM,Li2022PRB,Liu2021arXiv} leads to an enhancement of superconductivity, suggesting a competition relation between the CDW and superconductivity. In addition, the CDW order in \AVS, which has been demonstrated to be three-dimensional in nature~\cite{Liang2021PRX,Li2021PRX,Stahl2022PRB,Ortiz2021PRX,Xiao2022arXiv}, occurs in tandem with a giant anomalous Hall effect~\cite{Yang2020SA,Yu2021PRB} and rotation symmetry breaking~\cite{Xiang2021NC,Chen2021Nature,Li2022NP,Wu2022PRB}. While theoretical calculations have proposed a chiral flux phase with broken time-reversal symmetry for the CDW order in \AVS~\cite{Feng2021SB,Denner2021PRL}, which may account for the anomalous Hall effect, experimental evidence for time-reversal symmetry breaking remains controversial~\cite{Kenney2021JPCM,Jiang2021NM,Mielke2022Nature,Li2022PRBSTM,Li2022NP,Xu2022NP,Saykin2022arXiv}. Extensive studies suggest that the CDW in \AVS\ is an electronically driven charge order~\cite{Tan2021PRL,Li2021PRX}, where the saddle point or Fermi surface (FS) nesting plays a key role~\cite{Tan2021PRL,Denner2021PRL,Wang2021arXiv,Zhou2021PRB,Cho2021PRL,Nakayama2021PRB,Lou2022PRL}, but on the other hand, evidence for an electron-phonon coupling driven charge order has also been reported~\cite{Luo2022NC,Xie2022PRB,Si2022PRB,Liu2022NC}. To date, the nature of the CDW in \AVS\ and how it is interwoven with superconductivity remain elusive.

In general, an effective approach to studying the interplay between different orders is to chemically dope the material and then examine the doping evolution of the intertwined orders. Following this strategy, we investigate the transport and optical properties of the Nb-doped \CVNS\ (Nb-CVS) with $x = 0.03$ (Nb0.03) and 0.07 (Nb0.07), and compare them with those of the undoped \CVS\ (CVS). The Nb substitution suppresses $T_{\text{CDW}}$ and enhances $T_{c}$, consistent with the competition between the CDW and superconductivity. The Nb doping also leads to a drop of the residual resistivity ratio, pointing to a rise of the disorder level. The optical conductivity in the pristine phase can be well described by two Drude components (D1 and D2) for all compounds. However, the weight of D1 (D2) increases (decreases) with increasing $x$, which implies a change of the FS. The total weight of the Drude components is reduced as the Nb concentration grows, indicating an enhancement of electronic correlations. In the CDW state, while the CDW gap is clearly observed in the optical conductivity for all compounds, the gapped portion of the FS shrinks with increasing Nb content. Further analysis hints that the change of the FS, the enhancement of electronic correlations, the shrinkage of the gapped FS by the CDW order, and the increase of disorder may cooperatively act on the entanglement between the CDW and superconductivity in Nb-CVS.

%
%

\begin{figure}[tb]
\includegraphics[width=\columnwidth]{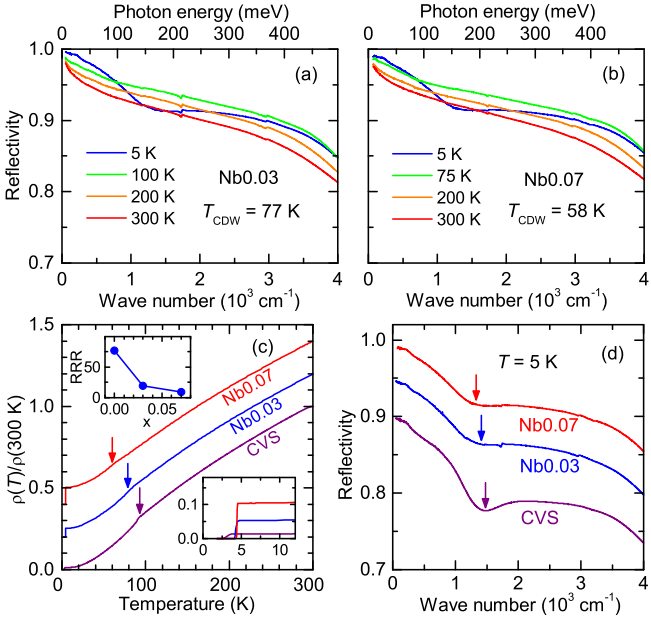}
\caption{(a) and (b) show $R(\omega)$ at different temperatures for Nb0.03 and Nb0.07, respectively. (c) The $T$-dependent resistivity $\rho(T)/\rho(300~\text{K})$ for CVS (purple solid curve), Nb0.03 (blue solid curve) and Nb0.07 (red solid curve). The curves for different compounds are shifted by 0.2 to avoid overlap. The top inset shows RRR as a function of $x$; the bottom inset displays $\rho(T)/\rho(300~\text{K})$ in the low-temperature range. (d) A comparison of $R(\omega)$ for CVS (purple solid curve), Nb0.03 (blue solid curve) and Nb0.07 (red solid curve) at 5~K. The spectra for different materials are shifted by 0.05 to avoid overlap.}
\label{CVNSRef}
\end{figure}

High-quality single crystals of \CVNS\ (Nb-CVS) with $x$ = 0.03 (Nb0.03) and 0.07 (Nb0.07) were synthesized using a self-flux method~\cite{Ortiz2019PRM}. Figure~\ref{CVNSRef}(c) shows the $T$-dependent resistivity $\rho(T)/\rho(300~\text{K})$ of CVS ($T_{\text{CDW}}$ = 92~K; $T_{c}$ = 3.08~K), Nb0.03 ($T_{\text{CDW}}$ = 77~K; $T_{c}$ = 4.2~K) and Nb0.07 ($T_{\text{CDW}}$ = 58~K; $T_{c}$ = 4.4~K). The arrow indicates the CDW transition for each compound. The bottom inset displays $\rho(T)/\rho(300~\text{K})$ in the low-temperature region, highlighting the superconducting transition. It is noticeable that the isovalent Nb substitution leads to a suppression of the CDW order and an enhancement of superconductivity. The top inset of Fig.~\ref{CVNSRef}(c) shows that the residual resistivity ratio (RRR) decreases with increasing Nb content, indicating a rise in the disorder level.

We measured the near-normal-incidence reflectivity $R(\omega)$ of Nb-CVS on a newly cleaved surface with the light polarized in the $ab$ planes using a Bruker Vertex 80V Fourier transform spectrometer. In order to obtain the absolute $R(\omega)$, we adopted an \emph{in situ} gold (or silver) evaporation technique~\cite{Homes1993AO}. Data in the frequency range of 50--12\,000~\icm\ were collected at numerous temperatures from 300 down to 5~K, followed by an extension of $R(\omega)$ to 50\,000~\icm\ at room temperature achieved through an AvaSpec-2048$\times$14 optical fiber spectrometer.

\begin{figure}[tb]
\includegraphics[width=\columnwidth]{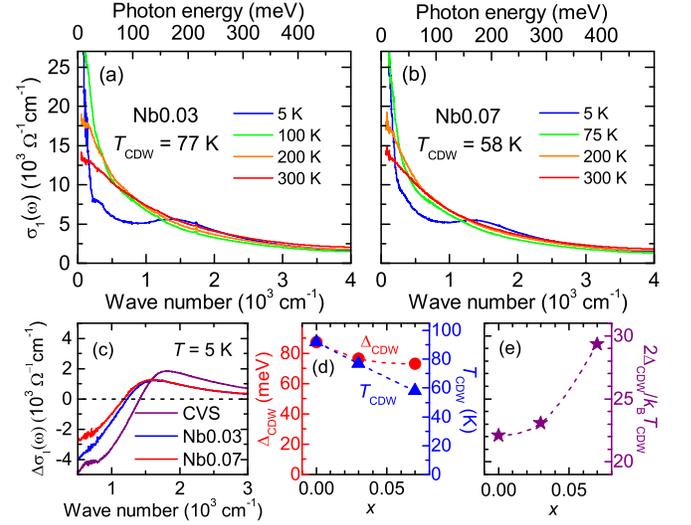}
\caption{(a) and (b) display $\sigma_{1}(\omega)$ at different temperatures for Nb0.03 and Nb0.07, respectively. (c) A comparison of $\Delta \sigma_{1}(\omega)$ for CVS (purple solid curve), Nb0.03 (blue solid curve) and Nb0.07 (red solid curve) at $T$ = 5~K. (d) $T_{\text{CDW}}$ and $\Delta_{\text{CDW}}$ as a function of $x$. (e) The evolution of $2\Delta_{\text{CDW}}/k_{\text{B}}T_{\text{CDW}}$ with $x$.}
\label{CVNSS1}
\end{figure}
%

\begin{figure*}[tb]
\includegraphics[width=\textwidth]{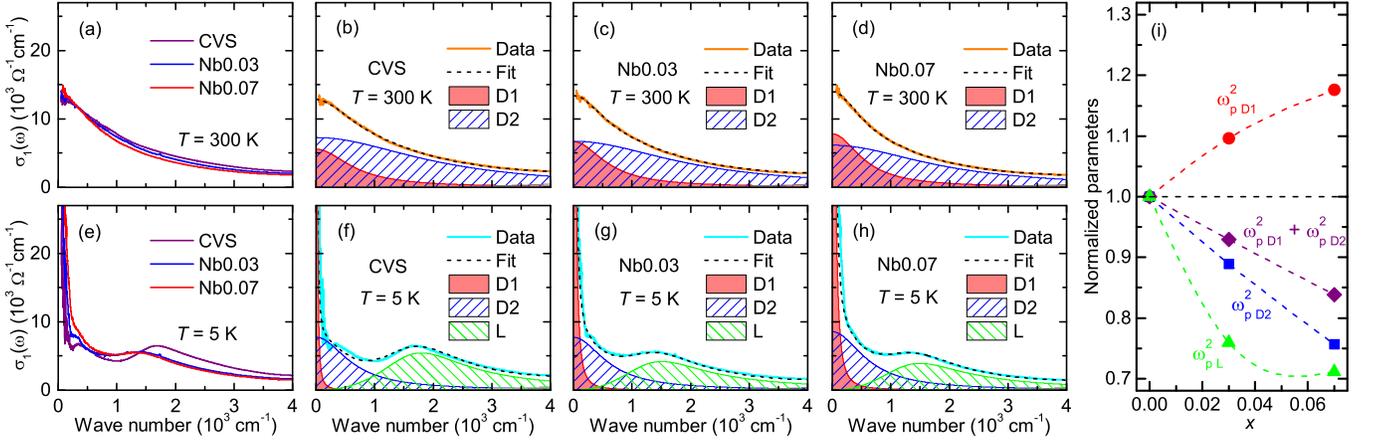}
\caption{(a) A comparison of $\sigma_{1}(\omega)$ for CVS, Nb0.03, and Nb0.07 at 300~K. (b)--(d) The fitting results at 300~K for CVS, Nb0.03, and Nb0.07, respectively. In each panel, the orange solid curve denotes the measured $\sigma_{1}(\omega)$ at 300~K; the black dashed line through the data is the fit, which is decomposed into a narrow Drude (D1, red shaded area) and a broad Drude (D2, blue hatched area) component. (e) compares $\sigma_{1}(\omega)$ for CVS, Nb0.03, and Nb0.07 at 5~K. (f)--(h) show the fitting results at 5~K for CVS, Nb0.03, and Nb0.07, respectively. In addition to D1 and D2, an extra Lorentz component (L, green hatched area) is introduced to describe the CDW gap feature. (i) The parameters extracted from the fit as a function of Nb substitution $x$. All parameters have been normalized to the values for CVS.}
\label{CVNSFit}
\end{figure*}

Figures~\ref{CVNSRef}(a) and \ref{CVNSRef}(b) show $R(\omega)$ up to 4000~\icm\ at several representative temperatures for Nb0.03 and Nb0.07, respectively. For both compounds, above $T_{\text{CDW}}$, such as $T$ = 300~K (red solid curve), the low-frequency $R(\omega)$ is very high and exhibits a Hagen-Rubens [$(1-R) \propto \sqrt{\omega}$] response, which is the characteristic of a metal; below $T_{\text{CDW}}$, for example $T$ = 5~K (blue solid curve), a pronounced suppression in $R(\omega)$ occurs at about 1300~\icm, signifying the opening of the CDW gap. The behavior of $R(\omega)$ in Nb-CVS closely resembles that in the undoped \AVS\ compounds~\cite{Zhou2021PRB,Zhou2022arXiv}. Nevertheless, as shown in Fig.~\ref{CVNSRef}(d), a comparison of $R(\omega)$ at 5~K for CVS (purple solid curve), Nb0.03 (blue solid curve), and Nb0.07 (red solid curve) reveals that the CDW-induced suppression in $R(\omega)$ becomes weaker and shifts to lower frequency with increasing Nb content, hinting at a weaker FS modification and a smaller CDW gap in Nb-CVS.

To gain more straightforward information, we obtained the real part of the optical conductivity $\sigma_{1}(\omega)$ through a Kramers-Kronig analysis of the measured $R(\omega)$~\cite{Dressel2002}. For both Nb0.03 and Nb0.07, as depicted in Figs.~\ref{CVNSS1}(a) and \ref{CVNSS1}(b), respectively, the low-frequency $\sigma_{1}(\omega)$ at $T > T_{\text{CDW}}$, e.g. $T$ = 300~K (red solid curve), is characterized by a conspicuous Drude response, consistent with the metallic nature of these materials. Below $T_{\text{CDW}}$, such as $T$ = 5~K (blue solid curve), the opening of the CDW gap is evidenced by a dramatic suppression of $\sigma_{1}(\omega)$ below about 1200~\icm\ alongside a recovery of the lost spectral weight [the area under the $\sigma_{1}(\omega)$ spectrum] at higher frequencies~\cite{Degiorgi1996PRL,Zhu2002PRB,Zhou2021PRB,Zhou2022arXiv}. The CDW gap value $\Delta_{\text{CDW}}$ can be extracted from the difference optical conductivity,
%
%
\begin{equation}
\Delta\sigma_{1}(\omega) = \sigma^{T < T_{\text{CDW}}}_{1}(\omega) - \sigma^{\text{P}}_{1}(\omega),
\label{Deltasigma}
\end{equation}
where $\sigma^{T < T_{\text{CDW}}}_{1}(\omega)$ and $\sigma^{\text{P}}_{1}(\omega)$ represent $\sigma_{1}(\omega)$ at temperatures below $T_{\text{CDW}}$ and that in the pristine phase, respectively. Figure~\ref{CVNSS1}(c) displays $\Delta \sigma_{1}(\omega)$ at $T$ = 5~K for CVS (purple solid curve), Nb0.03 (blue solid curve), and Nb0.07 (red solid curve), in which the zero-crossing point corresponds to $2\Delta_{\text{CDW}}$. It is obvious that $\Delta_{\text{CDW}}$ diminishes with increasing Nb content. Figure~\ref{CVNSS1}(d) plots the doping dependence of $T_{\text{CDW}}$ (blue solid triangles) and $\Delta_{\text{CDW}}$ (red solid circles). While both are suppressed by the Nb substitution, the suppression of $T_{\text{CDW}}$ is greater than that of $\Delta_{\text{CDW}}$. This effect gives rise to an enhancement of $2\Delta_{\text{CDW}}/k_{\text{B}}T_{\text{CDW}}$ with increasing Nb content, as traced out in Fig.~\ref{CVNSS1}(e).

%
%

We next examine the effects of Nb doping on $\sigma_{1}(\omega)$. As shown in Fig.~\ref{CVNSFit}(a), while the low-frequency $\sigma_{1}(\omega)$ spectra for all three compounds are dominated by a Drude-like response, the weight of the Drude peak is suppressed with increasing Nb doping. In order to quantitatively investigate the doping dependence of the optical properties in Nb-CVS, we fit the measured $\sigma_{1}(\omega)$ to the Drude-Lorentz model
%
%
\begin{equation}
\sigma_{1}(\omega) = \frac{2\pi}{Z_{0}} [
   \sum_{k} \frac{\omega^{2}_{p,k}}{\tau_{k}(\omega^{2}+\tau_{k}^{-2})} +
   \sum_{j} \frac{\gamma_{j} \omega^{2} \omega_{p,j}^{2}}{(\omega_{0,j}^{2} - \omega^{2})^{2} + \gamma_{j}^{2} \omega^{2}}],
\label{DLModel}
\end{equation}
where $Z_{0} \simeq 377$~$\Omega$ is the vacuum impedance. The first term is a sum of Drude components, which describe the optical response of free carriers or intraband transitions. Each Drude term has a plasma frequency $\omega_{p}$ and a quasiparticle scattering rate $1/\tau$. Here, $\omega_{p}^{2}$ and $1/\tau$ correspond to the spectral weight and the width of the Drude profile, respectively. The second term is a sum of Lorentzian oscillators, which are used to model localized carriers or interband transitions; $\omega_{0}$, $\gamma$, and $\omega_{p}$ refer to the resonance frequency, linewidth, and plasma frequency (strength), respectively.

For all three compounds [Figs.~\ref{CVNSFit}(b)--\ref{CVNSFit}(d)], the low-frequency $\sigma_{1}(\omega)$ at 300~K (orange solid curve) can be well described by the superposition (black dashed line) of a narrow Drude (D1, red shaded area) and a broad Drude (D2, blue hatched area) component. Recent optical studies on \AVS\ revealed a Drude component, a strong localization peak which has been assigned to a displaced Drude response, and multiple absorption bands arising from interband transitions in the far-infrared range~\cite{Uykur2021PRB,Uykur2022NPJQM,Wenzel2022PRB}. In contrast, the optical spectra of our \AVS~\cite{Zhou2022arXiv} and Nb-CVS [Figs.~\ref{CVNSFit}(c) and \ref{CVNSFit}(d)] samples only exhibit a plain Drude profile in the same spectral range. No localization peak or extra absorption bands are observed, suggesting that the Drude response is not displaced in our samples, and the low-energy interband transitions are much weaker. This discrepancy is likely to originate from the difference in \EF\ between samples grown by different groups, because an upward shift of \EF\ has been demonstrated to significantly enhance low-energy interband transitions~\cite{Uykur2021PRB,Uykur2022NPJQM,Wenzel2022PRB}. We notice that \EF\ in our samples~\cite{Nakayama2021PRB,Nakayama2022PRX,Kato2022PRL} is much lower than that from the calculations in Refs.~\cite{Uykur2021PRB,Uykur2022NPJQM,Wenzel2022PRB}, accounting for the absence of observable low-energy interband transitions in the optical spectra of our samples. Furthermore, we would like to stress that the two-Drude fit we employed here and in our previous work~\cite{Zhou2021PRB} does not conflict with the Drude plus displaced Drude model used in Refs.~\cite{Uykur2021PRB,Uykur2022NPJQM,Wenzel2022PRB}, as both the Drude and displaced Drude components are associated with intraband transitions, compatible with the multiband nature of these compounds. In our previous work on CVS~\cite{Zhou2021PRB}, D1 with a smaller $1/\tau$ has been assigned to the electron band near the $\Gamma$ point and the Dirac bands near the K point; D2, which features a larger $1/\tau$, has been associated with the saddle-point bands near the M point, because saddle points in the proximity of \EF\ act as scattering sinks that will give rise to strong quasiparticle scattering~\cite{Rice1975PRL}. The fit returns $\omega_{p}$ of the two Drude components for all compounds. Figure~\ref{CVNSFit}(i) depicts the evolution of $\omega_{p,D1}^{2}$ (red solid circles) and $\omega_{p,D2}^{2}$ (blue solid squares) with Nb concentration. $\omega_{p,D1}^{2}$ increases and $\omega_{p,D2}^{2}$ decreases with Nb doping, indicating an expansion of the FS formed by the electron band near $\Gamma$ or the Dirac bands near K and a shrinkage of the FS derived from the saddle-point bands near M. Recent first-principles calculations~\cite{Li2022PRB} and angle-resolved photoemission (ARPES)~\cite{Kato2022PRL} studies on Nb-CVS have revealed that the Nb substitution pushes the electron band at $\Gamma$ downward and simultaneously shifts the saddle-point bands near M upward, which leads to an expansion of the electron pocket at $\Gamma$ and a shrinkage of the FS formed by the saddle-point bands near M, in agreement with our optical results.

The total weight of the two Drude components $\omega_{p,D1}^{2} + \omega_{p,D2}^{2}$ [purple solid diamonds in Fig.~\ref{CVNSFit}(i)] decreases with increasing $x$, which correctly describes the effect observed from Fig.~\ref{CVNSFit}(a). Such a decrease in the total weight of the Drude components cannot be explained by the crystallographic disorder introduced by Nb doping, because disorder is expected to affect the width of the Drude profile which corresponds to the quasiparticle scattering rate $1/\tau$, but has no influence on the weight of the Drude profile, i.e. the integral of the Drude peak in $\sigma_{1}(\omega)$. The Drude weight is proportional to $n/m^{\ast}$, where $n$ and $m^{\ast}$ represent the carrier density and the effective mass, respectively. The decrease of the Drude weight may originate from a decrease of $n$, an increase of $m^{\ast}$, or the combination of both effects. As the isovalent substitution of Nb for V is not expected to cause effective carrier doping, the decrease of $\omega_{p,D1}^{2} + \omega_{p,D2}^{2}$ is likely to arise from an increase of $m^{\ast}$, which suggests an enhancement of electronic correlations in Nb-CVS. Such a change of electronic correlations induced by isovalent substitution has been observed in the 122 family of Fe-based superconductors BaFe$_{2}$(As$_{1-x}$P$_{x}$)$_{2}$~\cite{Nakajima2013PRB}. In BaFe$_{2}$(As$_{1-x}$P$_{x}$)$_{2}$, since the ionic radius of P is smaller than that of As, the substitution of P for As introduces chemical pressure which reduces the Fe-As/P bond length, thus increasing the overlap between the Fe and As/P atoms. As a consequence, the width of Fe bands is enlarged and the electronic correlations are reduced, resulting in an increase of the Drude weight with increasing P concentration. Here in CVS, although the V-$3d$ orbitals crossing \EF\ are subject to strong electronic correlations~\cite{Imada1998RMP}, the electronic correlations in this material have been found to be weak due to the hybridization between the V-$3d$ and the Sb-$p$ orbitals~\cite{Zhao2021PRB,Jeong2022PRB,Liu2022PRB,Sante2022arXiv}. The substitution of Nb for V expands the lattice parameters~\cite{Li2022PRB}, because the ionic radius of Nb is larger than that of V. The expansion of the lattice parameters reduces the hybridization between the V-$3d$ and Sb-$p$ orbitals, which is most likely to be responsible for the enhancement of electronic correlations in Nb-CVS.

Having examined the effect of Nb doping on $\sigma_{1}(\omega)$ in the pristine phase, we proceed to the analysis of $\sigma_{1}(\omega)$ in the CDW state. Figure~\ref{CVNSFit}(e) compares $\sigma_{1}(\omega)$ of CVS (purple solid curve), Nb0.03 (blue solid curve) and Nb0.07 (red solid curve) at 5~K. For all compounds, the low-frequency $\sigma_{1}(\omega)$ is suppressed due to the formation of the CDW gap. However, the suppression of the low-frequency $\sigma_{1}(\omega)$ becomes weaker and shifts to lower frequency upon Nb doping, indicating that the removed FS by the CDW gap shrinks and the value of $\Delta_{\text{CDW}}$ decreases. As shown in Figs.~\ref{CVNSFit}(f)--\ref{CVNSFit}(h), the measured $\sigma_{1}(\omega)$ at 5~K (cyan solid curve) for all three materials can also be well described by the Drude-Lorentz model (black dashed curve). In addition to D1 (red shaded area) and D2 (blue hatched area) which have been used for the fit in the pristine phase, an extra Lorentz component L (green hatched area) is required to describe the gap feature in $\sigma_{1}(\omega)$. While D1 becomes very narrow in width, the spectral weight of D2 is dramatically suppressed due to the opening of the CDW gap, and the removed spectral weight from D2 is transferred to L~\cite{Zhou2021PRB}. Hence, the weight of L, i.e. $\omega_{p,L}^{2}$, reflects the removed FS or density of states (DOS) near \EF\ by the CDW gap. Figure~\ref{CVNSFit}(i) shows that $\omega_{p,L}^{2}$ (green solid triangles) decreases with increasing Nb content, indicating that a smaller portion of the FS is removed by $\Delta_{\text{CDW}}$ in Nb-CVS.

%
The present experimental results may shed light on the interplay between the CDW and superconductivity. The evolution of $\sigma_{1}(\omega)$ with doping in Nb-CVS has revealed that the isovalent Nb substitution induces a change of the FS or electronic structure, an enhancement of electronic correlations, and a shrinkage of the removed FS by $\Delta_{\text{CDW}}$. Furthermore, the transport properties of Nb-CVS have proclaimed an increase of disorder caused by Nb substitution. All these effects may have a considerable impact on the CDW and superconductivity in this system. (i) Many studies have attested to the importance of the saddle point or FS nesting in stabilizing the CDW order in \AVS~\cite{Tan2021PRL,Denner2021PRL,Wang2021arXiv,Zhou2021PRB,Cho2021PRL,Nakayama2021PRB,Lou2022PRL}. The upward shift of the saddle-point bands near M induced by Nb doping moves the saddle point away from \EF, degrading the nesting of the saddle points or FS, which may be responsible for the suppression of the CDW order. In addition, recent work on Sn- and Ti-doped CVS suggests that superconductivity in CVS relies on the electron pocket derived from the Sb-$p_z$ orbitals near $\Gamma$~\cite{Liu2021arXiv,Oey2022PRM}. Therefore, the expansion of the electron pocket near $\Gamma$ may account for the enhancement of superconductivity in Nb-CVS. (ii) Previous theoretical calculations have documented that the ground state of a kagome system is sensitive to electronic correlations~\cite{Wang2013PRB,Kiesel2013PRL,Denner2021PRL,Ferrari2022PRB}. In transition metal dichalcogenides (TMDs), electronic correlations have been demonstrated to suppress the charge susceptibility or CDW order~\cite{Loon2018npjQM,Lin2020NC} but promote the spin susceptibility or spin fluctuations~\cite{Loon2018npjQM}, which are believed to mediate unconventional superconductivity~\cite{Moriya2000AP,Moriya2003RPP}. In this sense, the enhancement of electronic correlations induced by Nb doping in Nb-CVS may also play an important role in suppressing the CDW order and elevating $T_{c}$. Moreover, the larger $2\Delta_{\text{CDW}}/k_{\text{B}}T_{\text{CDW}}$ in Nb-CVS may also be related to the enhancement of electronic correlations. (iii) A theoretical model by Bilbro and McMillan~\cite{Bilbro1976PRB} shows that CDW order and superconductivity compete to open an energy gap on the common portion of the FS, which may occur in the \AVS\ system~\cite{Tan2021PRL,Qian2021PRB}. In Nb-CVS, as a smaller portion of the FS is removed by $\Delta_{\text{CDW}}$, more DOS near \EF\ is available for superconducting pairing, leading to the enhancement of $T_{c}$. (iv) Previous studies on TMDs~\cite{Cho2018NC,Li2017NPJQM,Mutka1983PRB} and cuprates~\cite{Leroux2019PNAS} have revealed that disorder suppresses $T_{\text{CDW}}$ by reducing the CDW correlation length and raises $T_{c}$, serving as an important tuning parameter for the competition between CDW and superconductivity. Here in Nb-CVS, the partial replacement of V with Nb inevitably introduces disorder to the V kagome planes. The increase of disorder has been evidenced by the decrease of RRR with increasing Nb content, as shown in the top inset of Fig.~\ref{CVNSRef}(c). Hence, the disorder arising from Nb substitution may also act as a key factor in manipulating the interplay between the CDW and superconductivity in Nb-CVS. Taking all the above facts into account, we deduce that the change of FS, the enhancement of electronic correlations, the shrinkage of CDW-gapped FS, and the increase of disorder may cooperatively manipulate the CDW and superconductivity in the \AVS\ family. The question of which effect plays a dominant role calls for further studies.

%
%
In summary, through examining the evolution of $\sigma_{1}(\omega)$ with Nb doping in Nb-CVS for both the pristine and the CDW phase, we found that: (i) The pristine-phase $\sigma_{1}(\omega)$ consists of two Drude components, denoted by D1 and D2, for all Nb-CVS and CVS compounds. D1 increases while D2 decreases in weight with Nb doping, suggesting a change of the FS. (ii) The total weight of the Drude components diminishes with increasing Nb content, which signifies an enhancement of electronic correlations. (iii) While the CDW gap is clearly resolved in $\sigma_{1}(\omega)$ for all compounds, the removed spectral weight by the CDW gap becomes smaller as the Nb content grows, indicating that a smaller portion of the FS is gapped. Considering these observations and the fact that Nb doping introduces disorder in the V kagome planes as evidenced by the decrease of RRR with $x$, a comprehensive analysis hints that the change of the FS, the enhancement of electronic correlations, the shrinkage of the CDW-gapped FS, and the increase of disorder may all play a role in manipulating the CDW order and superconductivity in Nb-CVS. Further studies are required to distinguish the dominant factor.

%
%

\begin{acknowledgments}
We thank Xinwei Fan, Huan Yang and Shunli Yu for helpful discussions. We acknowledge financial support from the National Natural Science Foundation of China (Grants No. 12174180, 11874206, 12061131001, 92065109, 11734003 and 11904294), the National Key R\&D Program of China (Grant No. 2020YFA0308800), Jiangsu shuangchuang program, the Beijing Natural Science Foundation (Grants No. Z210006, Z190006). Z.W. thanks the Analysis \& Testing Center at BIT for assistance in facility support.
\end{acknowledgments}

%

\end{document}